\documentclass[12pt,1p, authoryear]{elsarticle}
\usepackage{style}

\DeclareTextSymbol{\degr}{T1}{6}

\newcommand{\removed}[1]{}

\makeatletter
 \renewcommand\@journal{Materials Today Communication}
\makeatother
\begin{document}
%%%%%%%%%%%%%%%%%%%%%%%%%%%%%%%%%%%%%%%%%%%%%%%%%%%%%%%%%%%%%%%%%%%%%%%%%%%%%%%%%%%%%%%
% FRONT MATTER
%%%%%%%%%%%%%%%%%%%%%%%%%%%%%%%%%%%%%%%%%%%%%%%%%%%%%%%%%%%%%%%%%%%%%%%%%%%%%%%%%%%%%%%
\begin{frontmatter}
\title{Dependency of the Young's modulus to plastic strain in DP steels: a consequence of heterogeneity ?}
\author[Annecy]{Ludovic Charleux}
\author[Annecy]{Laurent Tabourot}
\author[Annecy]{Emile Roux}
\author[Annecy,Djibouti]{Moustapha Issack Farah}
\author[Annecy]{Laurent Bizet}
\address[Annecy]{Univ. Savoie Mont-Blanc, EA 4144, SYMME, F-74940, Annecy-le-Vieux, France}
\address[Djibouti]{Univ. of Djibouti, Djibouti}
\cortext[cor1]{Corresponding author: ludovic.charleux@univ-smb.fr}

%%%%%%%%%%%%%%%%%%%%%%%%%%%%%%%%%%%%%%%%%%%%%%%%%%%%%%%%%%%%%%%%%%%%%%%%%%%%%%%%%%%%%%%
% ABSTRACT
%%%%%%%%%%%%%%%%%%%%%%%%%%%%%%%%%%%%%%%%%%%%%%%%%%%%%%%%%%%%%%%%%%%%%%%%%%%%%%%%%%%%%%%
\begin{abstract}

The accurate springback prediction of dual phase (DP) steels has been reported as a major challenge. 
It was demonstrated that this was due to the lack of understanding of their nonlinear unloading behavior and especially the dependency of their unloading moduli on the plastic prestrain.
An improved compartmentalized finite element model was developed.
In this model, each element was assigned a unique linear elastic J2 plastic behavior without hardening.
The model's novelty lied in the fact that:
\begin{enumerate}[i)]
    \item a statistical distribution was discretized in a deterministic way and used to assign yield stresses to structures called compartments,
    \item those compartments were randomly associated with the elements through a random compartment element mapping (CEM).
\end{enumerate}
Multiple CEM were simulated in parallel to investigate the intrinsic randomness of the model. 
The model was confronted with experimental data extracted from the literature and it was demonstrated that the model was able to reproduce the dependence of the apparent moduli on the tensile prestrain.
It was also observed that the evolution of the apparent moduli was predicted even if it was not an explicit input of the experimental dataset used to identify the input parameters of the model. 
It was then deduced that the shape of the hardening and the dependency of moduli on the prestrain were two manifestations of a single cause: the heterogeneous yield stress in DP steels.

\end{abstract}

%%%%%%%%%%%%%%%%%%%%%%%%%%%%%%%%%%%%%%%%%%%%%%%%%%%%%%%%%%%%%%%%%%%%%%%%%%%%%%%%%%%%%%%
% KEYWORDS
%%%%%%%%%%%%%%%%%%%%%%%%%%%%%%%%%%%%%%%%%%%%%%%%%%%%%%%%%%%%%%%%%%%%%%%%%%%%%%%%%%%%%%%
\end{frontmatter}
%%%%%%%%%%%%%%%%%%%%%%%%%%%%%%%%%%%%%%%%%%%%%%%%%%%%%%%%%%%%%%%%%%%%%%%%%%%%%%%%%%%%%%%
% INTRODUCTION
%%%%%%%%%%%%%%%%%%%%%%%%%%%%%%%%%%%%%%%%%%%%%%%%%%%%%%%%%%%%%%%%%%%%%%%%%%%%%%%%%%%%%%%
\setcounter{footnote}{0}
\section{Introduction}

Dual phase (DP) steels exhibit an outstanding combination of strength and ductility.
They are widely used in the automotive industry where they contribute to the vehicle mass reduction and thus to greater fuel efficiency \citep{tasan_overview_2015}. 
This being said, the complexity of their mechanical behavior is especially high and many scientific challenges have to be addressed before they can be used with full background knowledge.
Among those, the accurate prediction of springback is one of the greatest \citep{wagoner_advanced_2013}.
It is known that the apparent Young's modulus of most metallic alloys is influenced by an applied prestrain \citep{Yoshida2002, pham_mechanical_2015} and that this is especially true in the case of DP steels.
It has also been demonstrated that this phenomenon affects the ability to predict springback \citep{eggertsen_constitutive_2010, Yu2009, ul2016springback}.
While the dependency of the apparent modulus to the prestrain is observed by many researchers, the mechanisms at stake are not yet fully determined. 
Therefore, several ways have been followed to take this evolution of the apparent modulus into account. 
The most widely used and arguably the most successful is the phenomenological modeling developed in numerous papers \citep{Yoshida2003, Kim2013, Sun2011, Lee2013, Ghaei2015a, Xue2016, Zajkani2017, torkabadi2018nonlinear}.
These studies do not focus on the underlying causes of the phenomenon but only aim to reproduce it, in most cases by using the model proposed by \citet{Yoshida2002}.
The drawback of phenomenological methods is that they rely on a fine-tuning stage of an ever-increasing number of parameters.

In parallel, other methods have been aiming at establishing a relation between the material's physical properties and its macroscopic behavior.
As pointed out by \citet{Paul2013}, the intrinsic dual phase heterogeneity of DP steels triggers strain incompatibility between the soft ferrite phase and the harder martensite phase. 
Several two-phase models have been developed using local phenomenological models as well as crystal plasticity to take into account the bimodal mechanical behaviors \citep{kadkhodapour_micro_2011,Ramazani2012,Moeini2017}.
Random microstructures have also been used with relative success \citep{Furushima2009, Furushima2013, Furushima2013b, Ayatollahi2016, khan2018microstructure}.
Still, this global simulation process has been unable to reproduce correctly the non-linear unloading behavior of DP steels and so it has been failing to determine the quantitative changes in the apparent modulus unless finally using the phenomenological Yoshida-Uemori model.

As stated by \citet{tasan_overview_2015}, the causes of heterogeneity in DP steels are multiple: on the one hand, heterogeneous dislocation microstructure, grain size distribution, presence of impurities and on the other hand, strong differentiation between the mechanical behaviors of its constituent phases which are themselves randomly distributed in the material.
Then, a third way has been explored to deal with these observations. 
It relies on introducing heterogeneity in the model in a more generalized way using a statistical spatial distribution of yield stresses \citep{Tabourot2012, Tabourot2013}.  
This compartmentalized model has shown that it can reproduce experimental observations with fewer adjustable parameters than phenomenological models and that their predictions are more realistic \citep{Bizet2017}.

In this paper, an improved version of the compartmentalized model was used to simulate and analyze the apparent modulus of DP steels. Section 2 is dedicated to the description of the compartmentalized model. Then, the model is applied to retrieve experimental data extracted from the literature. In section 3, the predictions of the model are discussed and confronted with the other existing modeling paradigms.

%%%%%%%%%%%%%%%%%%%%%%%%%%%%%%%%%%%%%%%%%%%%%%%%%%%%%%%%%%%%%%%%%%%%%%%%%%%%%%%%%%%%%%%
% MODEL
%%%%%%%%%%%%%%%%%%%%%%%%%%%%%%%%%%%%%%%%%%%%%%%%%%%%%%%%%%%%%%%%%%%%%%%%%%%%%%%%%%%%%%%
\section{The compartmentalized model }

A compartmentalized model is a random heterogeneous finite element model in which the material properties of every element arise from those of a substructure called a compartment.   
Each compartment has unique material properties. 
In the presented model each element is a compartment.
Compartments are not designed to represent a specific physical structure or scale such as grains. 
Their only purpose is to introduce a controlled amount of heterogeneity in the model to produce specific effects on the macroscopic mechanical behavior. 
The implementation of the compartmentalized model described in this section is made available by the authors (python libraries : \href{https://github.com/lcharleux/compmod2}{Compmod2} \citep{compmod2} and \href{https://github.com/lcharleux/argiope}{Argiope} \citep{argiope}, full code example : \href{https://compmod2.readthedocs.io/en/latest/notebooks_rst/DP_moduli/DP_steel.html}{Compdmod2 Documentation} \citep{compmod2-doc})

\subsection{Mesh and boundary conditions}

%%%%%%%%%%%%%%%%%%%%%%%%%%%%%%%%%%%%%%%%%%%%%%%%%%%%%%%%%%%%%%%%%%%%%%%%%%%%%%%%%%%%%%%
% FIGURE 0
%%%%%%%%%%%%%%%%%%%%%%%%%%%%%%%%%%%%%%%%%%%%%%%%%%%%%%%%%%%%%%%%%%%%%%%%%%%%%%%%%%%%%%%

\begin{figure*}
\begin{center}
\includegraphics[width = .6\textwidth]{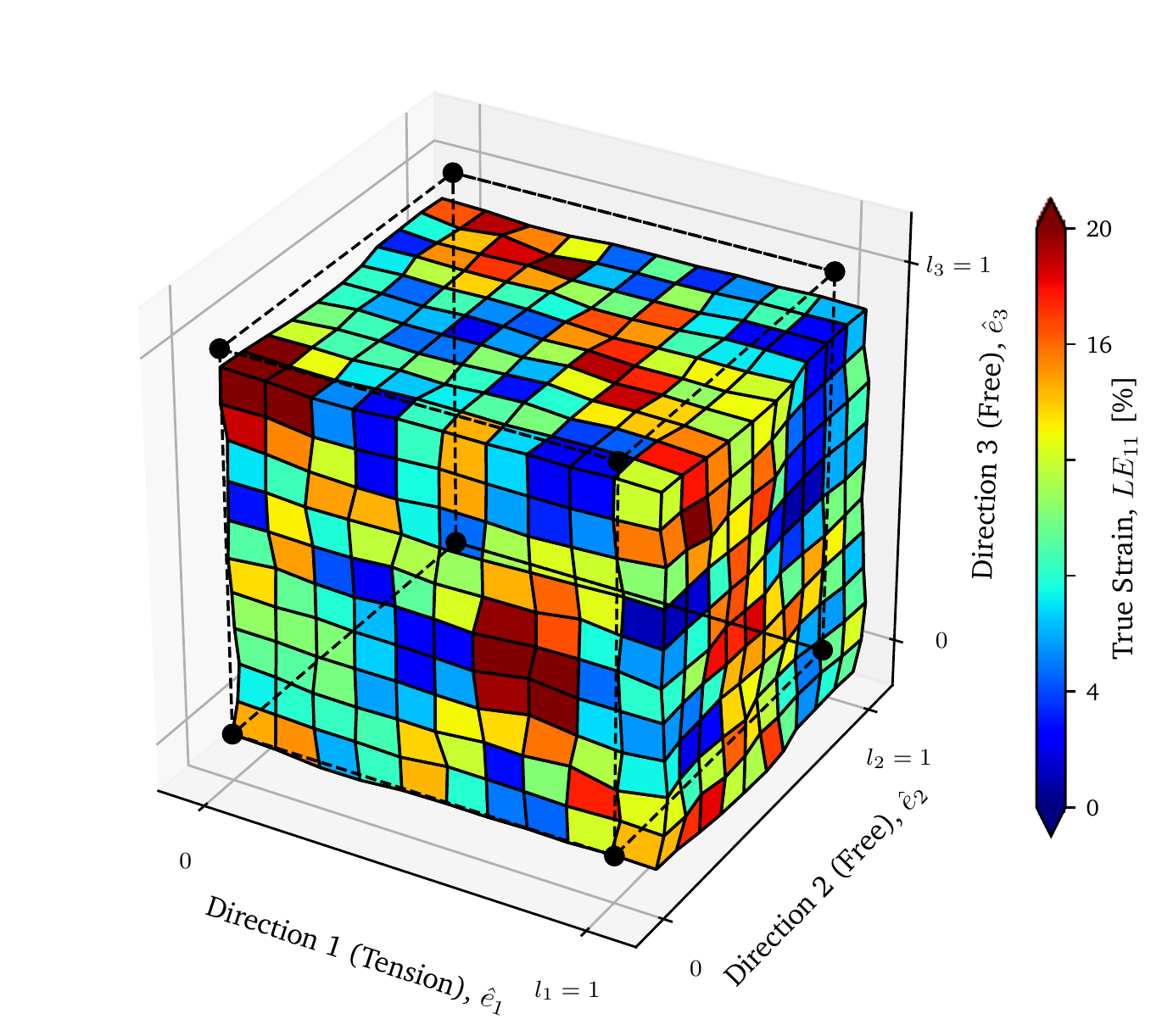}
\end{center}
\caption{The RVE is represented in a deformed state under macroscopic tensile strain. The edges of the cubic  RVE in its initial configuration are represented using dotted lines. The color map represents $LE_{11}$ local logarithmic total strain along the tested direction. The heterogeneity of the strain field is visible and a shear band can be observed on the top face. 
The periodic boundary conditions are also visible as all opposite faces are identical modulo a translation.}
\label{fig:model_3D}
\end{figure*}

In this paper, the finite element simulations were carried out using the commercial implicit solver Abaqus/Standard (2018 version). 
Fig. \ref{fig:model_3D} represents the cubic Representative Volume Element (RVE) test sample.
The initial dimensions of the RVE were $l\times l \times l$ along $(\hat e_1, \hat e_2, \hat e_3)$ where $l=1$. 
Only intensive properties such as stress and strains were extracted from the model.
Consequently, the problem was dimensionless and the value of $l$ had no impact on the results.
A structured $10 \times 10 \times 10 $ hexahedric mesh was used.
Periodic boundary conditions were applied in a similar way to \citep{wu2014applying}.
The sample was loaded in tension and the true tensile stress $\sigma$ as well and true tensile strain $\varepsilon$ were calculated.

\subsection{Compartmentalized material definition}

%%%%%%%%%%%%%%%%%%%%%%%%%%%%%%%%%%%%%%%%%%%%%%%%%%%%%%%%%%%%%%%%%%%%%%%%%%%%%%%%%%%%%%%
% FIGURE 1
%%%%%%%%%%%%%%%%%%%%%%%%%%%%%%%%%%%%%%%%%%%%%%%%%%%%%%%%%%%%%%%%%%%%%%%%%%%%%%%%%%%%%%%

\begin{figure}
\begin{center}
\includegraphics[width = .8\columnwidth]{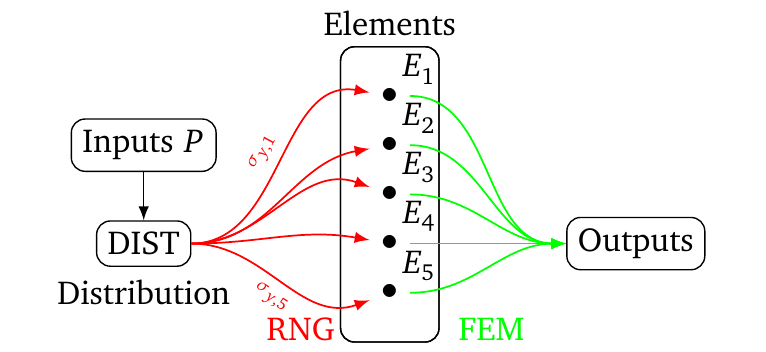}
\end{center}
\caption{Schematic representation of the intuitive implementation of the compartmentalized model. 
The input parameters $P$ are used to define the yield stress distribution DIST.  
A Random Number Generator (RNG) is used to calculate element yield stress $\sigma_{y,i}$ associated to each element $E_i$. 
The Finite Element Method is then used to solve the problem and generate outputs such as the tensile stress vs. tensile strain curve. 
The link between the input parameters and the output stress vs. strain curve is non-deterministic since two evaluations of the RNG with identical inputs lead to different outputs. 
Consequently, the optimization of the input parameters $P$ cannot be done with classical optimization algorithms.}
\label{fig:algorithm_old}
\end{figure}

\begin{figure}
\begin{center}
\includegraphics[width = 1.\columnwidth]{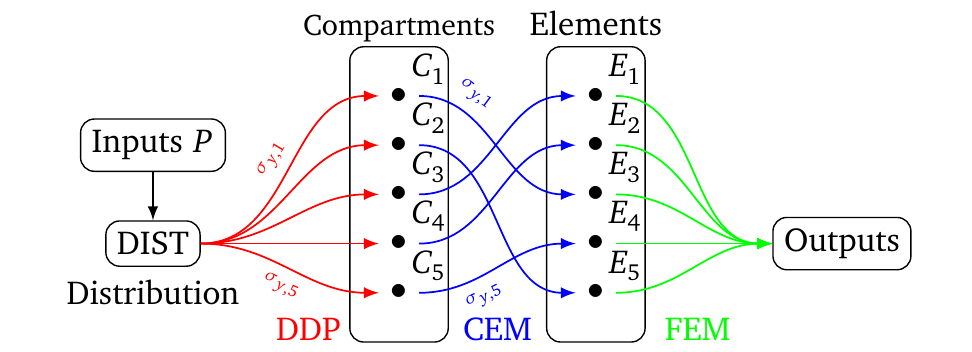}
\end{center}
\caption{Schematic representation of the new approach developed in this paper. 
As in previous approaches, the input parameters $P$ are used to define the yield stress distribution DIST. 
Then, the distribution is discretized using the Distribution Discretization Procedure (DDP) described in Fig. \ref{fig:distribution_discretization} and $N$ yield stress values $\sigma_{y,i}$ are calculated in the ascending order.
Each of them is associated with a compartment $C_i$.  
Subsequently, the Compartment Element Mapping (CEM) is applied to associate each compartment $C_i$ with a given element $E_j$. 
The CEM is initialized using a random draw algorithm. It can then be kept as is, for example during the input parameters optimization process. As a consequence, the model is deterministic, and therefore the optimization of the input parameters $P$ becomes possible with standard gradient-based algorithms such as Levenberg-Marquardt.}
\label{fig:algorithm_new}
\end{figure}

In a compartmentalized model, the material properties are distributed randomly among the elements of the mesh.  
This procedure has been greatly improved compared to the intuitive one used by the authors in previous articles \citep{Tabourot2012, Tabourot2013, Laurent2014, Bizet2017}.
This legacy procedure is described in Fig. \ref{fig:algorithm_old} while the new one is represented in Fig. \ref{fig:algorithm_new}.

%%%%%%%%%%%%%%%%%%%%%%%%%%%%%%%%%%%%%%%%%%%%%%%%%%%%%%%%%%%%%%%%%%%%%%%%%%%%%%%%%%%%%%%
%FIGURE 2
%%%%%%%%%%%%%%%%%%%%%%%%%%%%%%%%%%%%%%%%%%%%%%%%%%%%%%%%%%%%%%%%%%%%%%%%%%%%%%%%%%%%%%
  
Each compartment is associated with a unique but very basic, isotropic, linear elastic and J2 perfectly plastic material.
Consequently, no hardening is implemented in any of those materials.

Moreover, the elastic behavior of all compartments is homogeneous.
Their Young modulus has a fixed value $E = \unit{213}\giga\pascal$ and their Poisson's ratio is $\nu=0.3$. 
These values are in agreement with \cite{Chen2016b}. 

\subsection{Yield stress distribution}

The yield stresses $\sigma_y$ are distributed among the compartments following a statistical distribution noted DIST. 
It is defined by its probability density function (PDF) noted $f(\sigma_y)$ and its cumulative density function (CDF) noted $F(\sigma_y)$. 
In this case, the distribution is the weighted sum of two Weibull sub-distributions with PDFs $f_1(\sigma_y)$ and $f_2(\sigma_y)$ that verify:

\begin{equation}
f(\sigma_y) = w_1 f_1(\sigma_y) + (1-w_1) f_2(\sigma_y)
\end{equation}

And:

\begin{equation}
f_i(\sigma_y) = \dfrac{k_i}{\lambda_i} \left( \dfrac{\sigma_y}{\lambda_i} \right)^{k_i-1} \exp\left( -\left(\dfrac{\sigma_y}{ \lambda_i}\right)^{k_i}  \right)
\end{equation}

Where:
\begin{itemize}
    \item $k_i$ and $\lambda_i$ are respectively the shape parameters and the scale parameters of the Weibull sub-distributions,
    \item $w_1$ is the weighting factor between each sub-distribution and it verifies $w_1 \in \; ]0, 1[$.
\end{itemize}

\begin{figure*}
\begin{center}
\includegraphics[width = .8\columnwidth]{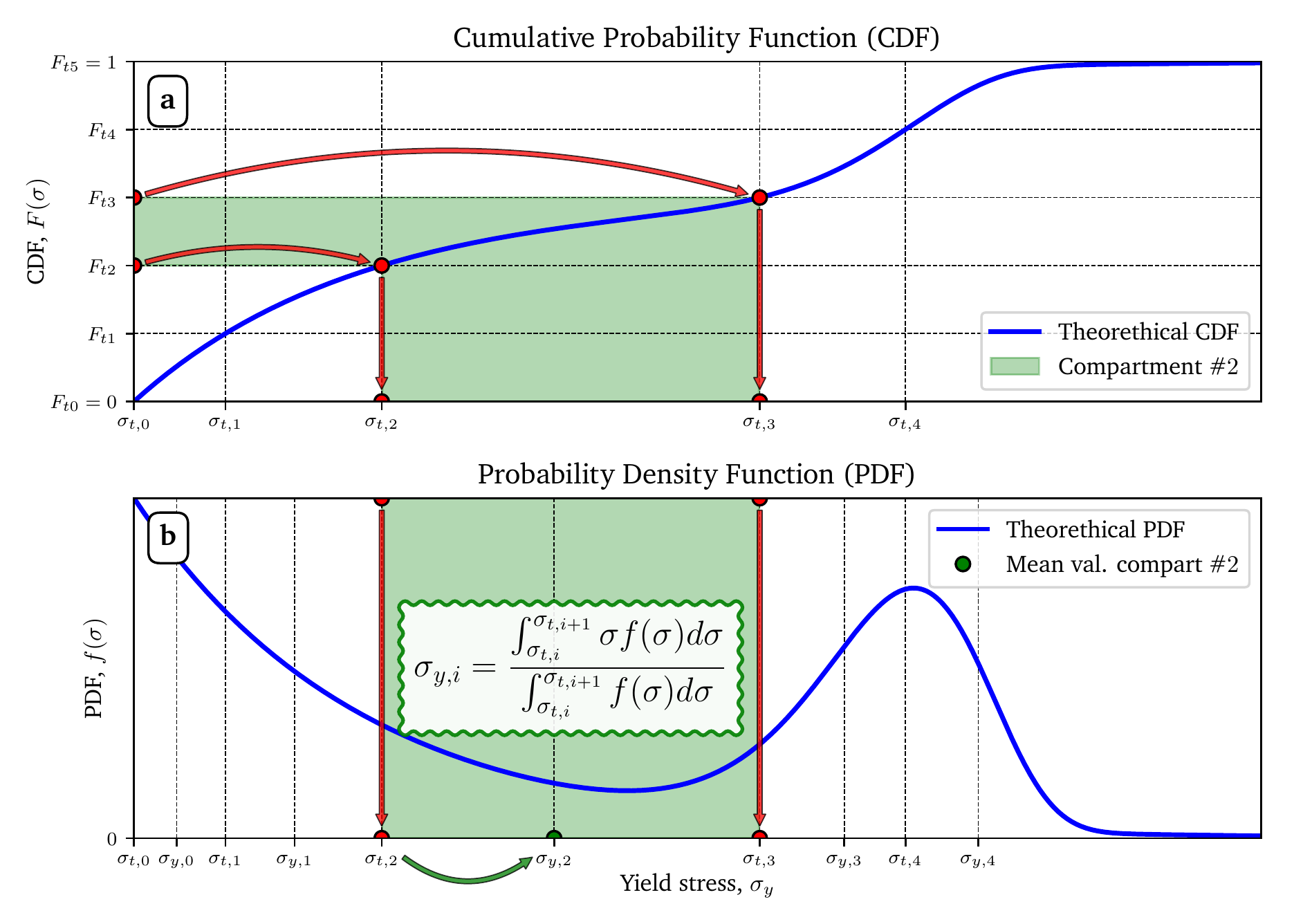}
\end{center}
\caption{The Distribution Discretization Procedure (DDP): \textbf{(a)} The CDF is split vertically into $N$ equal parts separated by $N+1$ threshold values $F_{t,i}$. The CDF is inversed  and stress threshold values $\sigma_{t,i}$ are determined. \textbf{(b)} The yield stress values $\sigma_{y,i}$ are calculated as the average value of the distribution's PDF $f$ over the $[\sigma_{t,i}, \sigma_{t_i+1}]$ interval. }
\label{fig:distribution_discretization}
\end{figure*}

The five input parameters $P = \left\lbrace k_1, k_2, \lambda_1, \lambda_2, w_1 \right \rbrace$ fully define the plastic behavior of the RVE. 
The values of the yield stresses associated with each compartment could then be calculated using a Random Number Generator (RNG) associated with the PDF defined above. 
However, this solution would mean that for a given value of $P$, multiple different sets of yield stress values could be possible because of the random nature of the RNG. 
As a consequence, the model would not be deterministic and most optimization scheme to identify the parameters would be compromised. 
To overcome this issue, a procedure has been developed to discretize the distribution in a deterministic way as described in Fig. \ref{fig:distribution_discretization}.  

%%%%%%%%%%%%%%%%%%%%%%%%%%%%%%%%%%%%%%%%%%%%%%%%%%%%%%%%%%%%%%%%%%%%%%%%%%%%%%%%%%%%%%%
%FIGURE 3
%%%%%%%%%%%%%%%%%%%%%%%%%%%%%%%%%%%%%%%%%%%%%%%%%%%%%%%%%%%%%%%%%%%%%%%%%%%%%%%%%%%%%%%

\begin{enumerate}
\item The model contains $N$ compartments, each occupying an equal part of the whole model volume.
As a consequence, each compartment $C_i$ is assumed to represent a cumulative probability of  $1/N$. 
The CDF is equally split along the vertical axis in $N$ parts separated by $N+1$ thresholds values noted $F_t$, where $F_{t0} = 0$ and $F_{tN}=1$.
\item The CDF is inverted using the Brent zero finding algorithm \citep{brent1973}  implemented in the Python library Scipy \citep{scipy}, $N+1$ threshold yield stress values $\sigma_t$ are determined. 
This process is described in Fig. \ref{fig:distribution_discretization}-a.
\item Each individual compartment then represents a unique portion of the distribution. Since only a single value of $\sigma_{y,i}$ has to be associated with each compartment $C_i$, the mean value of distribution on the interval $[\sigma_{t,i}, \sigma_{t,i+1}]$ is chosen:

\begin{equation}
\sigma_{y,i} = \dfrac{\int_{\sigma_{t,i}}^{\sigma_{t,i+1}} \sigma f(\sigma)d\sigma}{\int_{\sigma_{t,i}}^{\sigma_{t,i+1}} f(\sigma)d\sigma}
\end{equation}

\end{enumerate}

\subsection{The Compartment Element Mapping}

%%%%%%%%%%%%%%%%%%%%%%%%%%%%%%%%%%%%%%%%%%%%%%%%%%%%%%%%%%%%%%%%%%%%%%%%%%%%%%%%%%%%%%%
%FIGURE 4
%%%%%%%%%%%%%%%%%%%%%%%%%%%%%%%%%%%%%%%%%%%%%%%%%%%%%%%%%%%%%%%%%%%%%%%%%%%%%%%%%%%%%%%

\begin{figure*}
\begin{center}
\includegraphics[width=.8\textwidth]{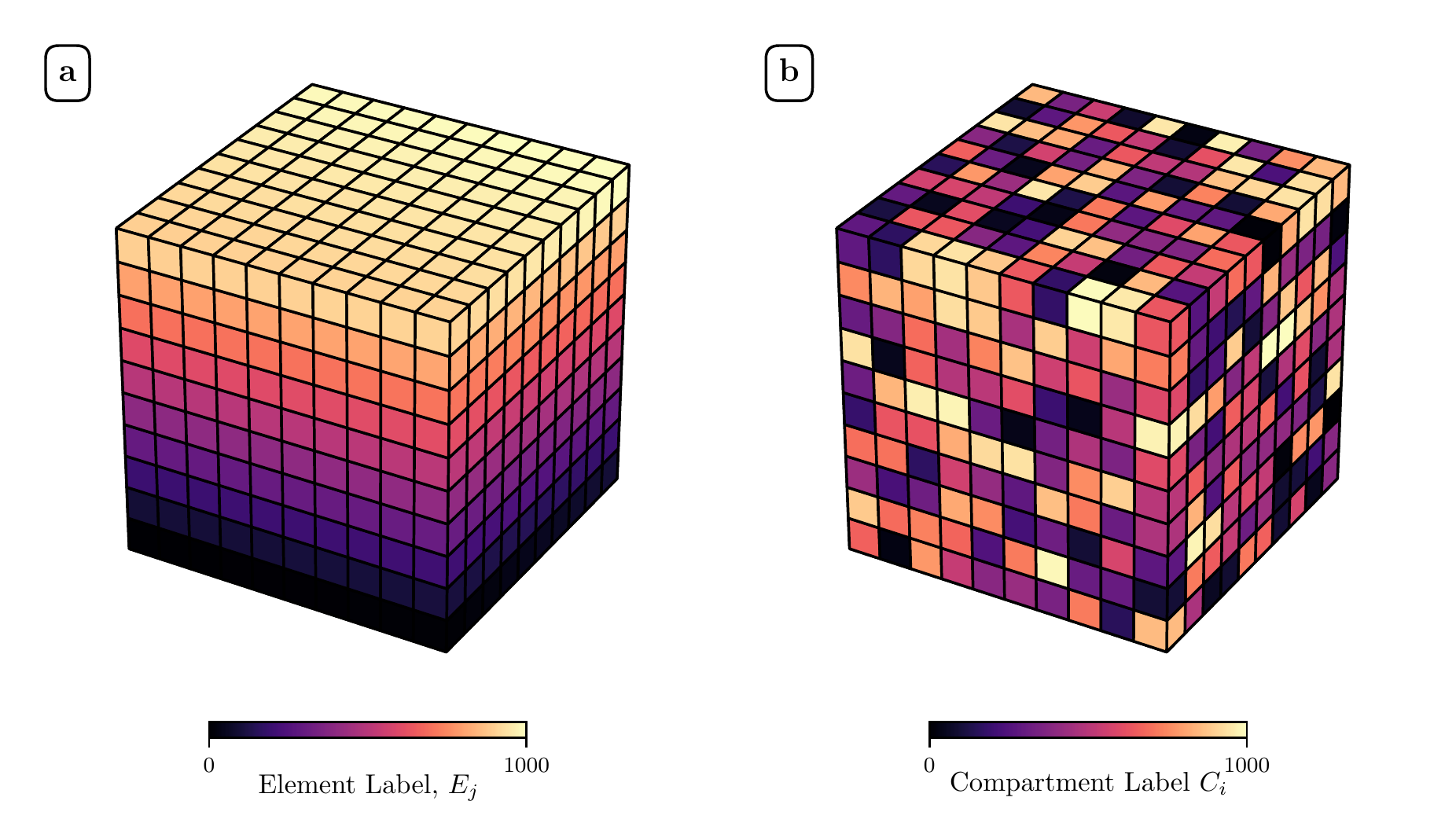}
\end{center}
\caption{Graphical illustration of the Compartment Element Mapping (CEM). \textbf{(a)} The elements are labeled in a standard way. \textbf{(b)} Each compartment is associated with one element by a bijective mapping. Each CEM instance is created by shuffling the element labels. 10 different CEM are used in the paper, each of them being referred to as CEM-0, \ldots, CEM-9. Only CEM-0 is represented here. }
\label{fig:CEM}
\end{figure*}

Each element $E_j$ is associated with a compartment $C_i$ by a discrete bijective mapping function referred to as the Compartment Element Mapping (CEM) as presented in Fig. \ref{fig:CEM}.
The CEM is initialized by shuffling the $\lbrace 1 , \ldots , N \rbrace$ list, creating the required association between the compartments and the elements. 
The CEM is the only source of the randomness of the model and thus it is the way to control it. 
As a consequence, as long as the CEM is not reinitialized, the model response to a given set of input parameters $P$ is deterministic.  
In this paper, a set of 10 random CEMs was generated by random shuffling and was used all along.

\subsection{Input parameter identification on a single loading-unloading-reloading cycle}

%%%%%%%%%%%%%%%%%%%%%%%%%%%%%%%%%%%%%%%%%%%%%%%%%%%%%%%%%%%%%%%%%%%%%%%%%%%%%%%%%%%%%%%
%FIGURE 5
%%%%%%%%%%%%%%%%%%%%%%%%%%%%%%%%%%%%%%%%%%%%%%%%%%%%%%%%%%%%%%%%%%%%%%%%%%%%%%%%%%%%%%%

\begin{figure*}
\begin{center}
\includegraphics[width = .8\textwidth]{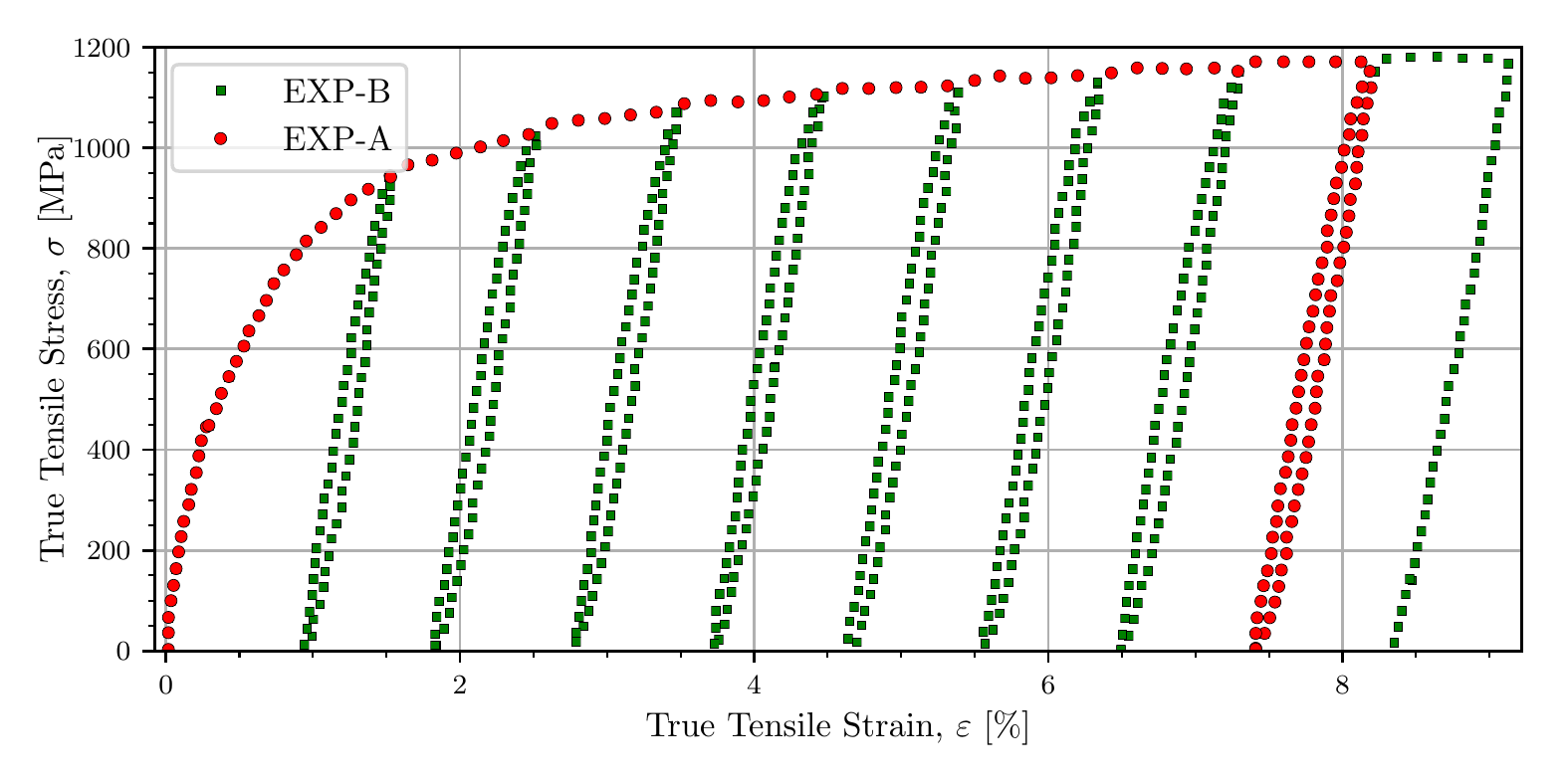}
\end{center}
\caption{The experimental dataset noted EXP is extracted from \cite{Ghaei2015} and is composed of multiple loading, unloading and reloading steps. The dataset is split into two sub-datasets: EXP-A contains all the loading steps and the last unloading-reloading steps whereas EXP-B contains the remaining data.}
\label{fig:exp_data}
\end{figure*}

The true tensile stress \textit{vs.} true tensile strain curve of a DP980 steel was extracted from Fig. 2 in \citet{Ghaei2015} and is referred to as the experimental curve. 
This curve was separated into multiple loading, unloading and reloading cycles (MUR).
The resulting data-set was split into two subsets represented in Fig. \ref{fig:exp_data}. 
The first subset noted \textbf{EXP-A} contains a monotonic loading up to 8\% of strain and the last unloading-reloading cycle (LUR).
It is essential to note that the evolution of the elastic moduli with the accumulated plastic strain cannot be observed on this subset because it contains only one unloading-reloading cycle. The second subset noted \textbf{EXP-B} contains the remaining experimental data.

For any given set of input parameters $P$ and for each CEM $c$, a true tensile stress response can be simulated following the strain path of the test \textbf{EXP-A}. 
As experimental and simulated data are not interpolated on the same grid, it is necessary to interpolate them.
The simulated stresses are therefore linearly interpolated on the strain value of the experimental curve \textbf{EXP-A}. 
The interpolated simulated stress is noted $\sigma_{ci}\left( P \right)$, where $i$ denote each point of the \textbf{EXP-A} curve.  
The stress residuals vector $\delta \sigma_{ci}\left( P \right)$ between the calculated set of stresses and the experimental stress $\sigma_{ei}$ is defined as follows:

\begin{equation}
\delta \sigma_{ci}\left( P \right) = \sigma_{ci} \left( P \right)- \sigma_{ei}   
\end{equation}

The residual vector $\delta\sigma_{ci}\left( P \right)$ gathers the point to point stress residual for all CEMs at each measurement points.

The optimal input parameters $P_{opt}$ were then calculated using the Levenberg-Marquardt least-square optimization algorithm \citep{leve44} by minimizing the residual vector $\delta\sigma_{ci}\left( P \right)$.

With an educated guess of the starting point of the Levenberg-Marquardt algorithm, the convergence was achieved after 70 evaluations of the cost function and thus after 700 individual simulations.
The optimal numerical values of $P_{opt}$ are given in the Table \ref{tab:params_opti}.

\begin{figure*}
\begin{center}
\includegraphics[width = .8\textwidth]{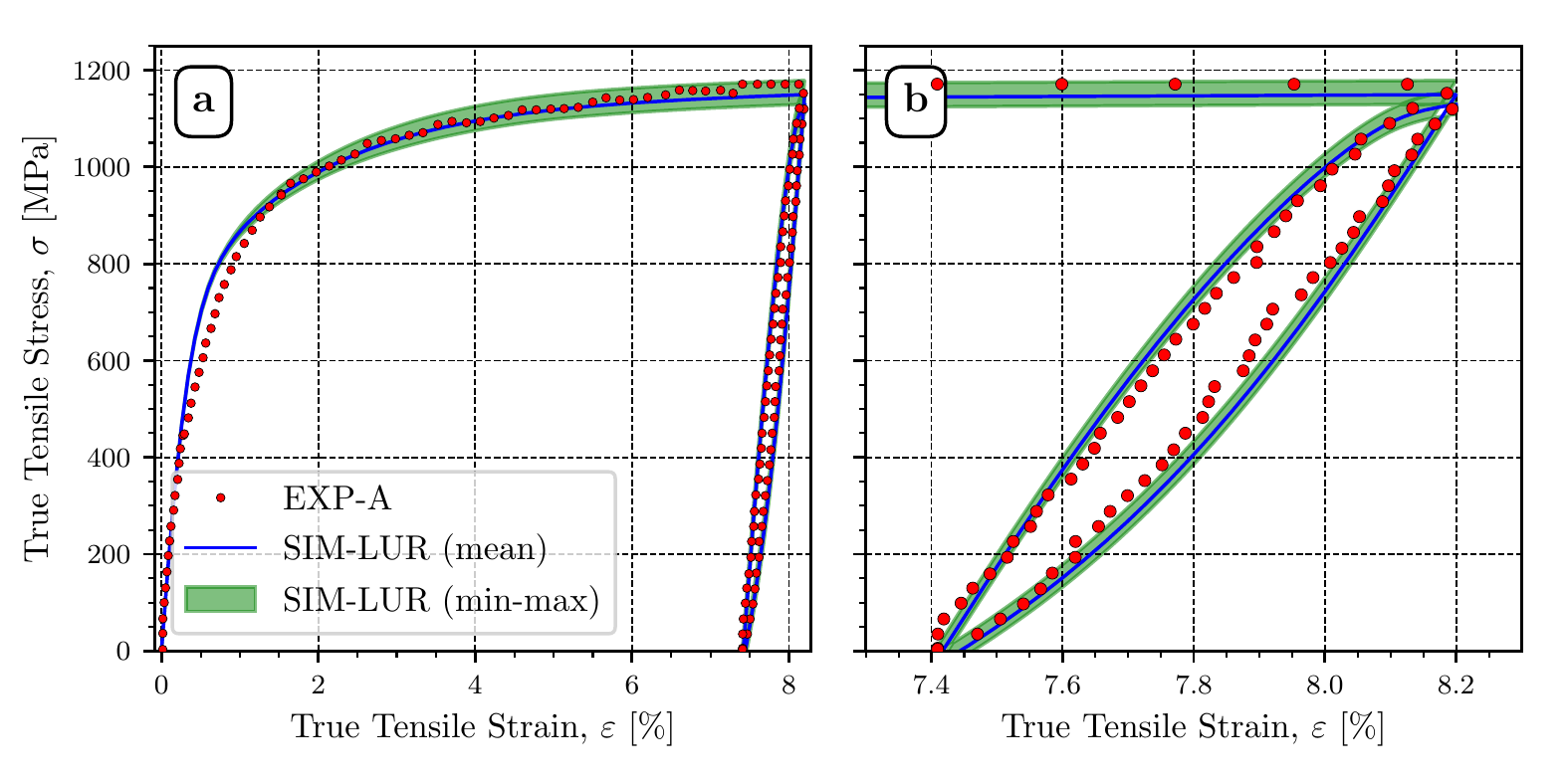}
\end{center}
\caption{The compartmentalized model is optimized by fitting the output true tensile stress vs. true tensile strain to the experimental sub dataset EXP-A. (a) The whole curve is represented. (b) Zoom on the unloading-reloading step.}
\label{fig:model_optimization}
\end{figure*}

\begin{figure*}
\begin{center}
\includegraphics[width = .8\textwidth]{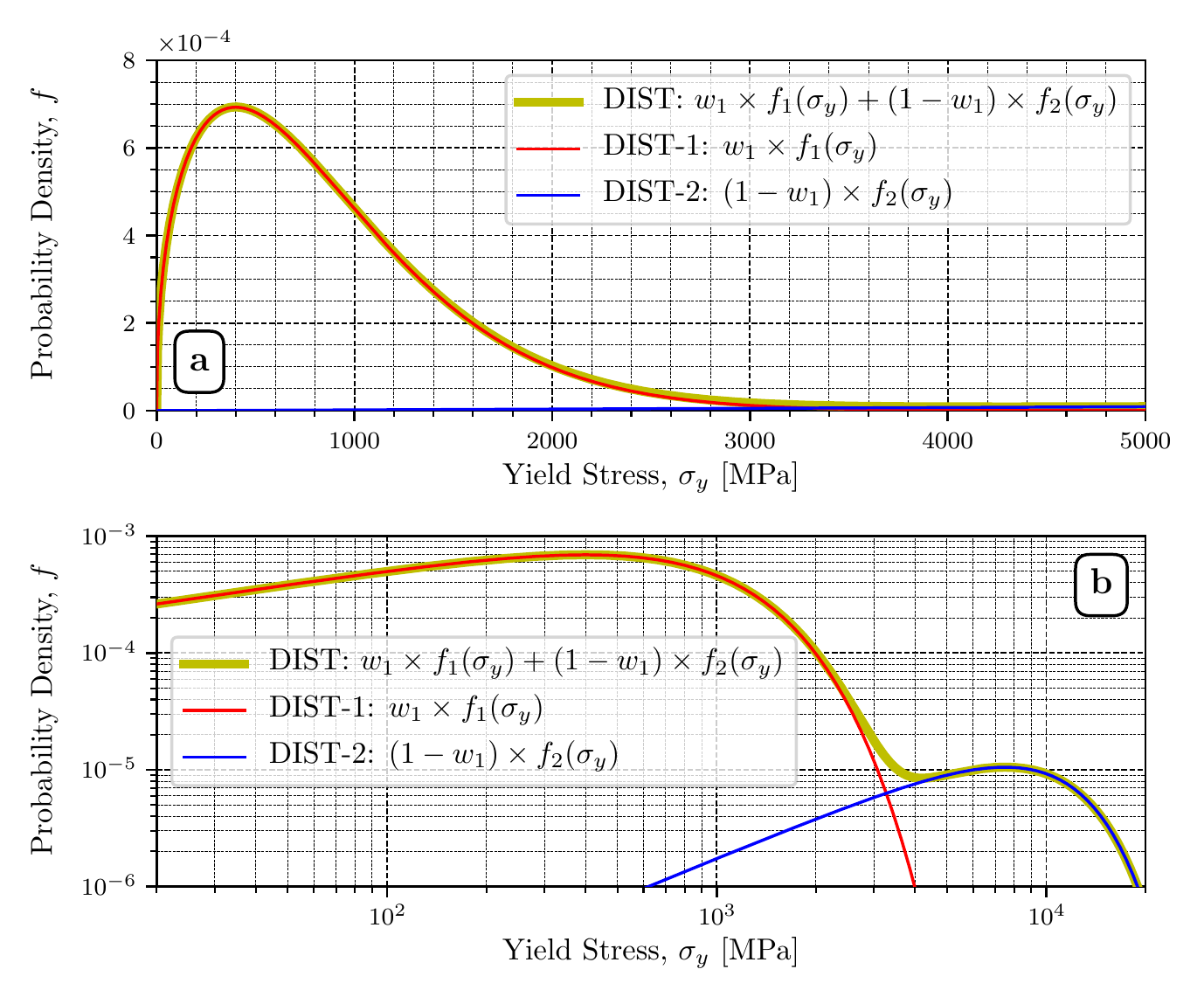}
\end{center}
\caption{ Representation of the statistical distribution of yield strength after optimization on the EXP-A dataset.
(a) linear scale representation. (b) log-log scale representation. The first Weibull distribution part DIST-1 is represented in red, the second DIST-2 is represented in blue, the total distribution DIST is represented in yellow. }
\label{fig:model_optimized_dist}
\end{figure*}

This way to proceed allowed us to determine not only the mean stress value but also the dispersion of the solutions associated with the different CEMs at each strain value. 
The mean stress value as well as min/max values are represented on Fig. \ref{fig:model_optimization}.
The associated yield stress distribution is represented in Fig. \ref{fig:model_optimized_dist}. 
The compartments yield stresses statistics are detailed in the Table \ref{tab:compartment_stats}.

%%%%%%%%%%%%%%%%%%%%%%%%%%%%%%%%%%%%%%%%%%%%%%%%%%%%%%%%%%%%%%%%%%%%%%%%%%%%%%%%%%%%%%%
%TAB 1
%%%%%%%%%%%%%%%%%%%%%%%%%%%%%%%%%%%%%%%%%%%%%%%%%%%%%%%%%%%%%%%%%%%%%%%%%%%%%%%%%%%%%%%

%%%%%%%%%%%%%%%%%%%%%%%%%%%%%%%%%%%%%%%%%%%%%%%%%%%%%%%%%%%%%%%%%%%%%%%%%%%%%%%%%%%%%%%
%FIGURE 6
%%%%%%%%%%%%%%%%%%%%%%%%%%%%%%%%%%%%%%%%%%%%%%%%%%%%%%%%%%%%%%%%%%%%%%%%%%%%%%%%%%%%%%%

%%%%%%%%%%%%%%%%%%%%%%%%%%%%%%%%%%%%%%%%%%%%%%%%%%%%%%%%%%%%%%%%%%%%%%%%%%%%%%%%%%%%%%%
% FIGURE 7
%%%%%%%%%%%%%%%%%%%%%%%%%%%%%%%%%%%%%%%%%%%%%%%%%%%%%%%%%%%%%%%%%%%%%%%%%%%%%%%%%%%%%%%

%%%%%%%%%%%%%%%%%%%%%%%%%%%%%%%%%%%%%%%%%%%%%%%%%%%%%%%%%%%%%%%%%%%%%%%%%%%%%%%%%%%%%%%
% TAB 2
%%%%%%%%%%%%%%%%%%%%%%%%%%%%%%%%%%%%%%%%%%%%%%%%%%%%%%%%%%%%%%%%%%%%%%%%%%%%%%%%%%%%%%%
\subsection{Moduli calculation on multiple loading-unloading-reloading cycles}

\begin{figure*}
\begin{center}
\includegraphics{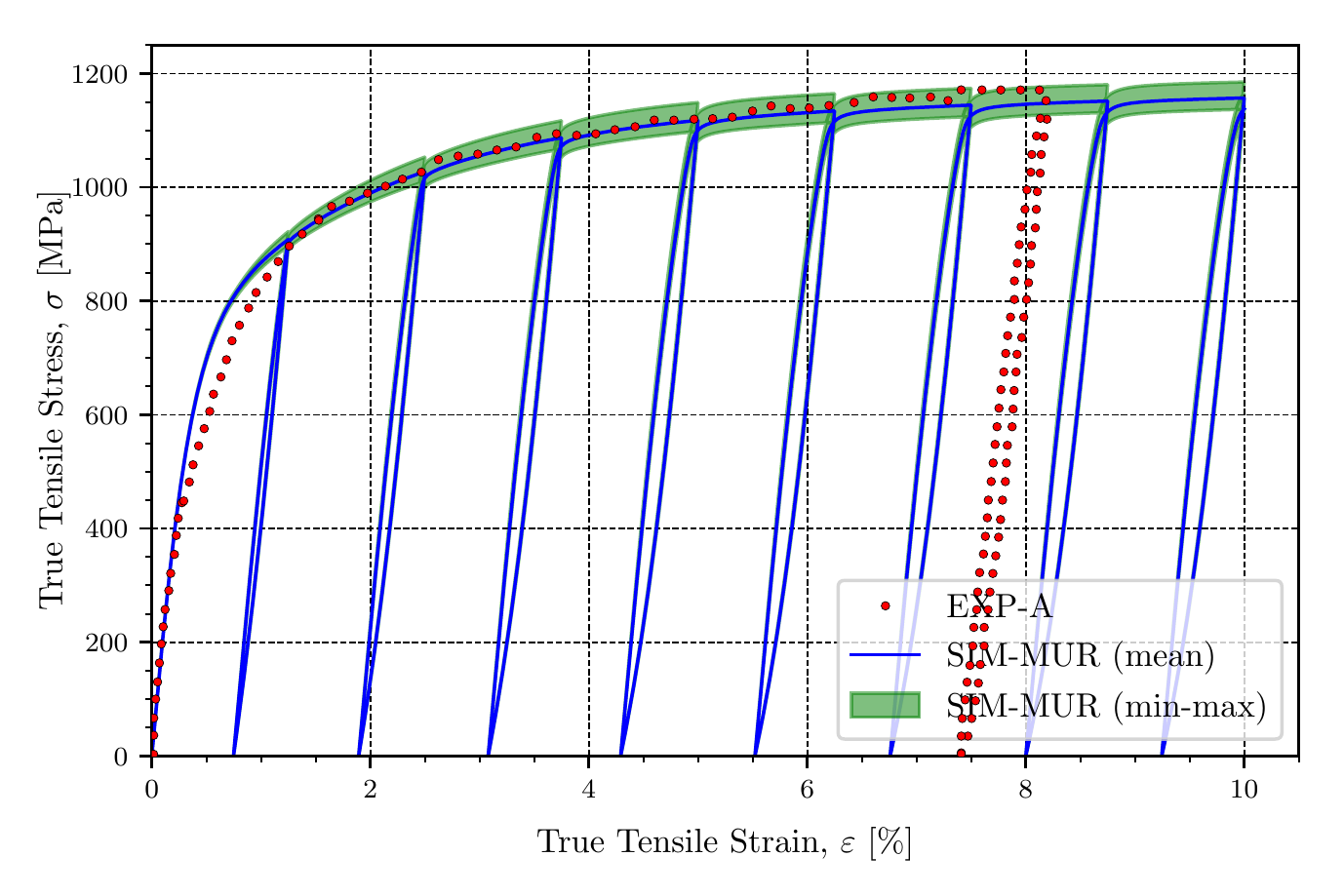}
\end{center}

\caption{The simulation called SIM-MUR using the optimal input parameters (identified on data subset EXP-A). It includes several loading-unloading-reloading cycles that allow to evaluate the evolution of the moduli according to the prestrain.} 
\label{fig:model_multi_cycle}
\end{figure*}

%%%%%%%%%%%%%%%%%%%%%%%%%%%%%%%%%%%%%%%%%%%%%%%%%%%%%%%%%%%%%%%%%%%%%%%%%%%%%%%%%%%%%%%
% FIGURE 8
%%%%%%%%%%%%%%%%%%%%%%%%%%%%%%%%%%%%%%%%%%%%%%%%%%%%%%%%%%%%%%%%%%%%%%%%%%%%%%%%%%%%%%%
A new set of 10 simulations were run using the optimized input parameter set $P_{opt}$. 
These new simulations included multiple unloading reloading cycles. 
It is important to note that these multiple unloading reloading cycles were not used in the parameters identification stage.
These multiple cycle simulations aimed to calculate the moduli as a function of the prestrain $\varepsilon_u$ to evaluate the capacity of the compartmentalized model. 
The obtained mean stress value as well as min/max values are represented on Fig. \ref{fig:model_multi_cycle}, and are noted \textbf{SIM-MUR} .

\begin{figure*}
\begin{center}
\includegraphics[width=.8\textwidth]{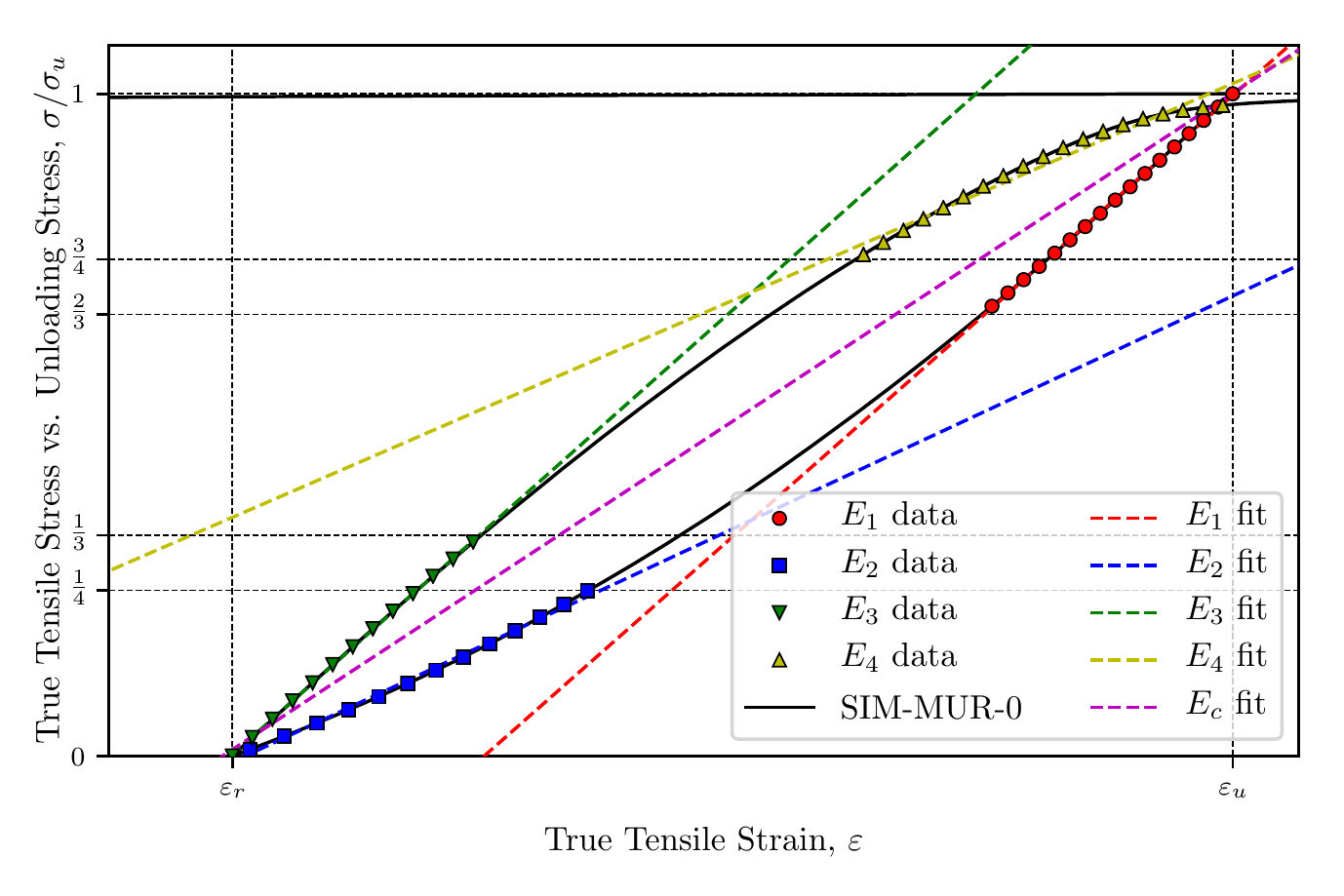}
\end{center}
\caption{
Schematical representation of the procedure introduced by \cite{Chen2016a} in order to calculate the 5 moduli $E_1$, $E_2$, $E_3$, $E_4$ and $E_c$ for each unloading reloading cycle. 
The vertical axis represents the normalized true stress $\sigma/\sigma_u$, where $\sigma_u$ is the true stress at the beginning of the unloading step. 
Each modulus is defined as the slope of a linear fit obtained over a given subset of the data points depending on which modulus is calculated.}
\label{fig:wagoner_moduli_definition}
\end{figure*}

The moduli $E_1$, $E_2$, $E_3$, $E_4$ and $E_c$ were then calculated following the definitions proposed by \cite{Chen2016a} as represented on Fig. \ref{fig:wagoner_moduli_definition} using the stress vs. strain curve \textbf{SIM-MUR}.

%%%%%%%%%%%%%%%%%%%%%%%%%%%%%%%%%%%%%%%%%%%%%%%%%%%%%%%%%%%%%%%%%%%%%%%%%%%%%%%%%%%%%%%
%FIGURE 9
%%%%%%%%%%%%%%%%%%%%%%%%%%%%%%%%%%%%%%%%%%%%%%%%%%%%%%%%%%%%%%%%%%%%%%%%%%%%%%%%%%%%%%%

%%%%%%%%%%%%%%%%%%%%%%%%%%%%%%%%%%%%%%%%%%%%%%%%%%%%%%%%%%%%%%%%%%%%%%%%%%%%%%%%%%%%%%%
% RESULTS AND DISCUSSION
%%%%%%%%%%%%%%%%%%%%%%%%%%%%%%%%%%%%%%%%%%%%%%%%%%%%%%%%%%%%%%%%%%%%%%%%%%%%%%%%%%%%%%%
\section{Results and discussion}

\subsection{Overall performances of the compartmentalized model}

The Fig. \ref{fig:model_optimization} demonstrates that the compartmentalized model can reproduce efficiently the experimental loading/unloading behavior of a DP steel (DP980 here). 
This is also true for other existing models such as the QPE model \citep{Sun2011}.
However, the compartmentalized model is more efficient as it only relies on a homogeneous linear elastic coupled to a J2 plastic criterion without hardening behavior and 5 additional parameters (Tab. \ref{tab:params_opti}) to describe the material.
This specificity makes the compartmentalized model's parameters easier to identify than those of its phenomenological counterparts. 
In this paper, only 70 optimization iterations (700 simulations) were needed to identify the parameters required to fit the \textbf{EXP-A} experimental sub-data-set.

\subsection{Compartment statistics}
\label{subsec:stats}

The identified parameters $P_{opt}$ solely describe the level of heterogeneity of the local yield stresses.
Fig. \ref{fig:model_optimized_dist} gives an accurate description of the optimal yield stress distribution.

Compared to other alloys previously modeled using the compartmentalized model \citep{Bizet2017}, it appears that distinguishing feature of the DP980 steel behavior is due to its exceptional heterogeneity that leads to a wide range of distributed yield stresses.
In that respect, 3 compartment sets are defined in the Tab. \ref{tab:compartment_stats}, each of them producing a given effect or property.
\begin{description}
\item[The Soft to Hard compartments (SHC)] constitute the main lobe of the distribution and represent a fraction of $80.5\% $. If this population has to be modeled alone, a single Rayleigh distribution could be used. 
The upper bound of the yield stress of SHC fixed at \unit{3000}\mega\pascal~because all the elements having higher yield stress never exhibited plastic strain in any simulation. 
This also means that the local von Mises stress field can sometimes reach very high values close to $\unit{3}\giga\pascal$.
\item[The Elastic compartments (ELC)] represent $12.3\%$ of the compartments. 
They are key compartments to model the specificity of materials exhibiting hardening at large strains such as DP980. 
If all compartments would become plastic during loading, the hardening would saturate. 
Hence, to exhibit hardening even at $\varepsilon \approx 10\%$, a significant proportion of the compartments must always remain elastic. 
\item[The Ultra-Soft compartments (USC)] represent $7.2\%$ of the compartments. 
They exhibit very low yield stresses. 
They don't play a key role during loading but their existence is strongly connected with the Bauschinger effect observed during unloading and the hysteretic behavior associated with unloading-reloading cycles. 
Indeed, at the end of the loading step (Fig. \ref{fig:model_optimization},  $\varepsilon \approx 8\%$ and $\sigma \approx \unit{1200}\mega\pascal$), when the unloading starts, most of these compartments rapidly become plastic in the compressive direction during the first $\unit{200}\mega\pascal$ of the unloading. 
Thus, these compartments dissipate energy and increase the curvature of the unloading stress vs. strain curve at the end of the unloading step. 
The same phenomenon appears symmetrically during reloading.
\end{description}

%%%%%%%%%%%%%%%%%%%%%%%%%%%%%%%%%%%%%%%%%%%%%%%%%%%%%%%%%%%%%%%%%%%%%%%%%%%%%%%%%%%%%%%
\subsection{Moduli evolution predicted by the compartmentalized model}

%%%%%%%%%%%%%%%%%%%%%%%%%%%%%%%%%%%%%%%%%%%%%%%%%%%%%%%%%%%%%%%%%%%%%%%%%%%%%%%%%%%%%%%
%FIGURE 10
%%%%%%%%%%%%%%%%%%%%%%%%%%%%%%%%%%%%%%%%%%%%%%%%%%%%%%%%%%%%%%%%%%%%%%%%%%%%%%%%%%%%%%%

\begin{figure*}
\begin{center}
\includegraphics[width=.8\textwidth]{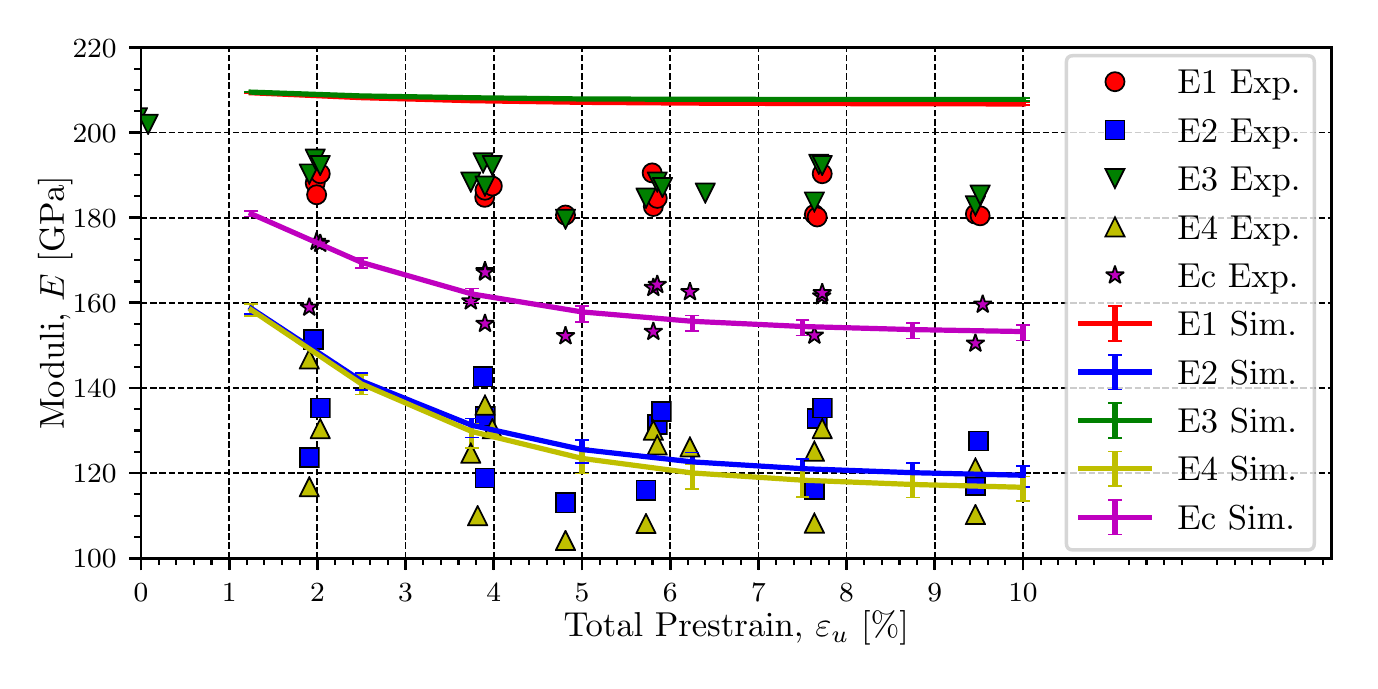}
\end{center}
\caption{The evolution of the 5 apparent moduli calculated on SIM-MUR is compared to the experimental data extracted from Fig. 5 of \citet{Chen2016a}.}
\label{fig:model_modulus_Wagoner}
\end{figure*}

Fig. \ref{fig:model_modulus_Wagoner} represents the experimental measurements taken from Fig. 5 in \citet{Chen2016a} as well as the values obtained by \textbf{SIM-MUR} using the protocol described in Fig. \ref{fig:wagoner_moduli_definition}.
Fig. \ref{fig:model_modulus_Wagoner} shows that the evolution of $E_1$ and $E_3$ is overestimated by the compartmentalized model. 
However, in their paper, \citet{Chen2016a} indicate that the decrease observed in their $E_1$ and $E_3$ measurements is small in the face of measurement uncertainties and could thus be an artifact related to a lack of resolution.
In contrast, the variations of $E_2$, $E_4$ and $E_c$ calculated by the simulation \textbf{SIM-MUR} using $P_{opt}$ are in very good agreement with the experimental data.
This implies, on the one hand, that the compartmentalized model is capable of reproducing the moduli decrease with a constant intrinsic Young's modulus $E$. On the other hand, since the optimization of the input parameters $P_{opt}$ is carried out with the \textbf{EXP-A} dataset, which contains only one loading/unloading/reloading cycle, it follows that, in the context of the compartmentalized model, the decay of the apparent moduli is related to the shape of this cycle.
These results indicate that strain-hardening, as well as the evolution of apparent moduli, are only two manifestations of a single cause: the heterogeneity of the material. It also shows that the compartmentalized model allows a physically realistic representation of the mechanisms involved in the plastic deformation of DP steels.

%%%%%%%%%%%%%%%%%%%%%%%%%%%%%%%%%%%%%%%%%%%%%%%%%%%%%%%%%%%%%%%%
% CONCLUSION
%%%%%%%%%%%%%%%%%%%%%%%%%%%%%%%%%%%%%%%%%%%%%%%%%%%%%%%%%%%%%%%%
\section{Conclusion}

In this work, the compartmentalized model has been proved to be a relevant way to model the behavior of DP steels. The model was improved by separating its intrinsic spatial heterogeneity and its randomness into two independent contributions driven respectively by the DDP and the CEM. It has been shown that this improvement allows the model to be deterministic and thus that its input parameters $P$ could be easily optimized as long as the CEM is kept unchanged. 
Results obtained from model simulations have been confronted with experimental pieces of evidence found in the literature and it has been demonstrated that:
\begin{enumerate}
    \item it can reproduce the overall shape of the stress vs. strain curve of the DP980 steel without needing the use of the variable intrinsic Young's modulus.
    \item the evolution of the apparent moduli with the level of prestrain is predicted spontaneously when the model is optimized to reproduce the experimental LUR cycle.
\end{enumerate}
Consequently, it has been postulated that the evolution of the moduli with the level of prestrain is just a consequence of the level of heterogeneity of mechanical properties such as yield stresses. 
The double Weibull yield stress distribution with perfect elastic-plastic behavior is sufficient to give a very good global strain hardening representation.

%%%%%%%%%%%%%%%%%%%%%%%%%%%%%%%%%%%%%%%%%%%%%%%%%%%%%%%%%%%%%%%%
% APPENDIX
%%%%%%%%%%%%%%%%%%%%%%%%%%%%%%%%%%%%%%%%%%%%%%%%%%%%%%%%%%%%%%%%
%\appendix

%%%%%%%%%%%%%%%%%%%%%%%%%%%%%%%%%%%%%%%%%%%%%%%%%%%%%%%%%%%%%%%%
% ACKNOWLEDGEMENTS
%%%%%%%%%%%%%%%%%%%%%%%%%%%%%%%%%%%%%%%%%%%%%%%%%%%%%%%%%%%%%%%%

\section*{Acknowledgements}
The authors would like to thank the French National Research Agency for its financial backing (XXS FORMING / ANR-12-RMNP-0009).

%%%%%%%%%%%%%%%%%%%%%%%%%%%%%%%%%%%%%%%%%%%%%%%%%%%%%%%%%%%%%%%%
% BIBLIOGRAPHY
%%%%%%%%%%%%%%%%%%%%%%%%%%%%%%%%%%%%%%%%%%%%%%%%%%%%%%%%%%%%%%%%

%\section*{References}
%\bibliographystyle{elsarticle-harv} 
%\bibliography{Hybrid-Models.bib}

\newpage

%%%%%%%%%%%%%%%%%%%%%%%%%%%%%%%%%%%%%%%%%%%%%%%%%%%%%%%%%%%%%%%%
% TABLES
%%%%%%%%%%%%%%%%%%%%%%%%%%%%%%%%%%%%%%%%%%%%%%%%%%%%%%%%%%%%%%%%
% TAB 1: PARAM OPTI
\begin{table}
\begin{center}
\begin{tabular}{cc} 
\hline
\textbf{Parameter} & \textbf{Value} \\
$k_1$ & $1.64$ \\
$k_2$ & $3.17$ \\
$l_1$ & $4.16\times 10^{-3}$ \\
$l_2$ & $3.42\times 10^{-2}$ \\  
$w_1$ & $8.67\times 10^{-1}$ \\
\hline
\end{tabular}
\end{center}
\caption{The parameters $P_{opt}$ resulting from the optimization for the compartmentalized model.}
\label{tab:params_opti}
\end{table}

% TAB 2: MODEL COMPARTMENT STATS
\begin{table*}
\begin{center}
\begin{tabular}{cccccc} 
\hline
\textbf{Label}                & $\sigma_{y,min}$        & $\sigma_{y,max}$ & $\bar \sigma_{y}$ & $N_c$ &   \textbf{Behavior}\\
USC & \unit{0}\mega\pascal    & \unit{200}\mega\pascal  & \unit{123}\mega\pascal   & 72 & Ultra Soft \\
SHC & \unit{200}\mega\pascal  & \unit{3000}\mega\pascal & \unit{865}\mega\pascal   & 805 & Soft to Hard \\
ELC & \unit{3000}\mega\pascal & $+\infty$               & \unit{6790}\mega\pascal   & 123 & Elastic \\
\hline
\end{tabular}
\end{center}
%\caption{Statistics on the compartment yield stresses associated with the optimized solution $P_{opt}$.}
\caption{Statistical yield stress distribution in the optimized solution $P_{opt}$.}
\label{tab:compartment_stats}
\end{table*}

%%%%%%%%%%%%%%%%%%%%%%%%%%%%%%%%%%%%%%%%%%%%%%%%%%%%%%%%%%%%%%%%
% FIGURES
%%%%%%%%%%%%%%%%%%%%%%%%%%%%%%%%%%%%%%%%%%%%%%%%%%%%%%%%%%%%%%%%
\newpage
% FIGURE 0: DEFORMED MESH

% FIGURE 1: ALGORITHM OLD VERSION

% FIGURE 2: ALGORITHM NEW VERSION

% FIGURE 3: DISTRIBUTION DISCRETIZATION

% FIGURE 4: CEM

% FIGURE 5: EXP DATA

% FIGURE 6: MODEL OPTIMIZATION

% FIGURE 7: MODEL OPTIMIZATION DISTRIBUTION

% FIGURE 8: MODEL MULTI CYCLE STRESS VS. STRAIN

% FIGURE 9: WAGONER MODULI DEFINITION

% FIGURE 10: WAGONER MODULI VALUES

%%%%%%%%%%%%%%%%%%%%%%%%%%%%%%%%%%%%%%%%%%%%%%%%%%%%%%%%%%%%%%%%%%%%%%%%%%%%%%%%%%%%%%%
\end{document}